\documentclass[12pt]{article}
\usepackage{epsfig} \usepackage{amssymb} \usepackage{amsfonts}
\usepackage{psfig}

\newcommand{\bea}{\begin{eqnarray}}
\newcommand{\eea}{\end{eqnarray}}

\newcommand{\as}{\alpha_s}
\newcommand{\asMZ}{\alpha_s(M^2_Z)}

\begin{document}

\begin{center}
{\Large \bf QCD coupling constant value and
deep inelastic measurements
} \\

\vspace{4mm}

V.G. Krivokhijine and A.V. Kotikov\\ 

\vspace{4mm}
Joint Institute for Nuclear
Research, 141980 Dubna, Russia
\end{center}

\begin{abstract}
We reanalyze 
deep inelastic scattering data 
of BCDMS Collaboration by including proper cuts of  ranges
with large systematic errors. 
We perform also the 
fits of high statistic deep inelastic scattering data 
of BCDMS, SLAC, NM and BFP Collaborations  
taking the data separately and in combined way and find good agreement
between these analyses. We
extract the values of
the QCD coupling constant $\alpha_s(M^2_Z)$ up to
NLO level.
\end{abstract}

\section{ Introduction }

The deep inelastic scattering (DIS) leptons on hadrons is the basical
 process to study the values of the parton distribution functions (PDF)
which are universal (after choosing of factorization and renormalization 
schemes) and
can be used in other processes.
The accuracy of the present data for deep inelastic
structure functions (SF) reached the level at which
the $Q^2$-dependence of logarithmic QCD-motivated terms and power-like ones
may be studied separately 
(for a review, see the recent papers \cite{Beneke} and references 
therein).

In the present letter we sketch the results of our analysis \cite{KriKo}
at the next-to-leading order (NLO)
of perturbative QCD for
the most known DIS SF 
$F_2(x,Q^2)$ 
\footnote{Here $Q^2=-q^2$ and $x=Q^2/(2pq)$ are standard DIS variables,
where $q$ and $p$ are photon and hadron momentums, respectively.}
taking into account experimental data \cite{SLAC1}-\cite{BFP} of
SLAC, NM,  BCDMS and BFP Collaborations.
We
stress the power-like effects, so-called twist-4 (i.e.
$\sim 1/Q^2$) 
contributions.
To our purposes we represent the SF $F_2(x,Q^2)$ as the contribution
of the leading twist part $F_2^{pQCD}(x,Q^2)$ 
described by perturbative QCD, 
when the target mass corrections are taken into account
(and coincides with $F_2^{tw2}(x,Q^2)$
when the target mass corrections are withdrawn), and the  
nonperturbative part (``dynamical'' twist-four terms):
\vskip -0.5cm
\begin{equation}
F_2(x,Q^2) 
\equiv F_2^{full}(x,Q^2)
=F_2^{pQCD}(x,Q^2)\,
\Bigl(
1+\frac{\tilde h_4(x)}{Q^2}
\Bigr),
\label{1}
\end{equation}
where $\tilde h_4(x)$ is magnitude of twist-four terms.

Contrary to standard fits (see, for example, \cite{Al2000}- \cite{fits}) 
when the direct
numerical calculations based on 
Dokshitzer-Gribov-Lipatov-Altarelli-Parisi 
(DGLAP) 
equation \cite{DGLAP} are used to evaluate structure functions, 
we use the exact solution of DGLAP equation
for the Mellin moments $M_n^{tw2}(Q^2)$ of
SF $F_2^{tw2}(x,Q^2)$:
\vskip -0.5cm
\begin{equation}
M_n^{k}(Q^2)=\int_0^1 x^{n-2}\,F_2^{k}(x,Q^2)\,dx~~~~~~~ (
k=full, pQCD, tw2, ...)
\label{2}
\end{equation}
\vskip -0.3cm
\noindent
and
the subsequent reproduction of $F_2^{k}(x,Q^2)$
at every needed $Q^2$-value with help of the Jacobi 
Polynomial expansion method \cite{Barker,Kri}
(see similar analyses at the NLO level 
\cite{Kri,Vovk}
and at the next-next-to-leading order (NNLO) level and above \cite{PKK}.

In this letter we 
do not present exact formulae of $Q^2$-dependence
of SF $F_2$ which are 
given in \cite{KriKo}. We note only that
the moments $M_{n}^{tw2}(Q^2)$ at 
some $Q^2_0$ is theoretical input of our analysis and 
the twist-four term $\tilde h_4(x)$
is considered as a set of free parameters (one constant
$\tilde h_4(x_i)$ per $x_i$-bin):
$\tilde h_4^{free}(x)=\sum_{i=1}^{I} \tilde h_4(x_i)$, 
where $I$ is the number of bins.

\vskip -0.3cm
\section{ 
Fits of $F_2$: procedure}
\label{sec:form}

\vskip -0.2cm
Having the QCD expressions for the Mellin moments
$M_n^{k}$ we can reconstruct the SF $F_2^k(x)$
as
\vskip -0.5cm
\begin{equation}
F_{2}^{k,N_{max}}(x,Q^2)=x^{a}(1-x)^{b}\sum_{n=0}^{N_{max}}
\Theta_n ^{a , b}
(x)\sum_{j=0}^{n}c_{j}^{(n)}{(\alpha ,\beta )}
M_{j+2}^{k} \left ( Q^{2}\right ),
\label{2.1}
\end{equation}
\vskip -0.3cm
\noindent
where $\Theta_n^{a,b}$ are the Jacobi polynomials
\footnote{We 
note here that there is similar method 
\cite{Ynd}, based on Bernstein polynomials. The method has been used 
in the analyses at the NLO level in \cite{KaKoYaF}
and at the NNLO level in \cite{SaYnd}.}
and $a,b$ are fitted parameters.

First of all, we choose the cut $Q^2 \geq 1$ GeV$^2$ in all our studies.
For $Q^2 < 1$ GeV$^2$, the applicability of twist expansion is very
questionable. 
Secondly, we
choose quite large values of the normalization point
$Q^2_0$: our
perturbative formulae should be applicable at the value of
$Q^2_0$. Moreover, the higher order corrections $\sim \as^k(Q^2_0)$ 
and $\sim (\as(Q^2)-\as(Q^2_0))^k$
($k \geq 2$) should be
less important at higher $Q^2_0$ values.

We use MINUIT program \cite{MINUIT} for
minimization of 
$\chi^2(F_2) = {\left|(F_2^{exp} - F_2^{teor})/\Delta F_2^{exp}
\right| }^2$. 
We consider
free normalizations of data for different experiments. 
For the reference, we use the most stable deuterium BCDMS data
at the value of energy $E_0=200$ GeV 
($E_0$ is the initial energy lepton beam). 
Using other types of data as reference gives
negligible changes in our results. The usage of fixed normalization
for all data leads to fits with a bit worser $\chi^2$.

\vspace{-0.3cm}
\section{ Results of fits }

\vskip -0.2cm
Hereafter 
we choose
 $Q^2_0$ = 90 GeV$^2$ ($Q^2_0$ = 20 GeV$^2$) for the 
nonsinglet (combine nonsinglet and singlet) evolution, 
 that is in good agreement with above 
conditions. We use also $N_{max} =8$.

\vspace{-0.2cm}
\subsection { BCDMS ~~${}^{12}C + H_2 + D_2$ data }

We start our analysis with the most precise experimental data 
\cite{BCDMS1} obtained  by BCDMS muon
scattering experiment at the high $Q^2$ values.
The full set of data is 762 (607) points (for the bounded 
$x$ range: $x \geq 0.25$).

It is well known that the original analyses 
given by BCDMS Collaboration itself (see
also Ref. \cite{ViMi}) lead to quite small values 
 $\alpha_s(M^2_Z)=0.113$.
Although in some recent papers (see, for example, 
\cite{Al2000,H1BCDMS})
more higher values of the coupling constant 
$\alpha_s(M^2_Z)$ have been observed, we think that
an additional reanalysis of BCDMS data should be very useful. 

Based on study \cite{Kri2} 
we proposed in
\cite{KriKo} that 
the reason for small values
of $\alpha_s(M^2_Z)$ coming from BCDMS data was the existence of the subset
of the data having large systematic errors. 
We studied this subject by 
introducing several so-called $Y$-cuts 
\footnote{Hereafter we use the kinematical variable $Y=(E_0-E)/E_0$,
where 
$E$ is
scattering energies of lepton.} 
(see \cite{KriKo}). 
Excluding this set of data with large systematic errors
leads to essentially larger values of $\alpha_s(M^2_Z)$ and very slow
dependence of the values on the concrete choice of the $Y$-cut (see below).

We use the following $x$-dependent $Y$-cuts:
\vskip -0.7cm
\bea
& &y \geq 0.14 \,\,~~~\mbox{ when }~~~ 0.3 < x \leq 0.4, ~~~~~~~ 
y \geq 0.16 \,~~~\mbox{ when }~~~ 0.4 < x \leq 0.5 \nonumber \\
& &y \geq Y_{cut3} ~~~\mbox{ when }~~~ 0.5 < x \leq 0.6, ~~~~~~
y \geq Y_{cut4} ~~~\mbox{ when }~~~ 0.6 < x \leq 0.7 \nonumber \\
& &y \geq Y_{cut5} ~~~\mbox{ when }~~~ 0.7 < x \leq 0.8 
\label{cut}
\eea
\vskip -0.3cm
\noindent
and
several $N$ sets  for the cuts at $0.5 < x \leq 0.8$: 
%
\begin{table}[h]
\begin{center}
\begin{tabular}{|c|c|c|c|c|c|c|c|}
\hline
$N$ & 0 & 1 & 2 & 3 & 4 & 5 & 6 \\
\hline \hline
$Y_{cut3}$ & 0 & 0.14 & 0.16 & 0.16 & 0.18 & 0.22 & 0.23 \\  
$Y_{cut4}$ & 0 & 0.16 & 0.18 & 0.20 & 0.20 & 0.23 & 0.24 \\
$Y_{cut5}$ & 0 & 0.20 & 0.20 & 0.22 & 0.22 & 0.24 & 0.25 \\
\hline
\end{tabular}
\caption{The values of $Y_{cut3}$, $Y_{cut4}$ and $Y_{cut5}$.
}\label{tab2}
\end{center}
\end{table}

\begin{figure}[t]
\begin{minipage}[t]{0.48\linewidth}
\begin{center}
\includegraphics[width=3.1in]{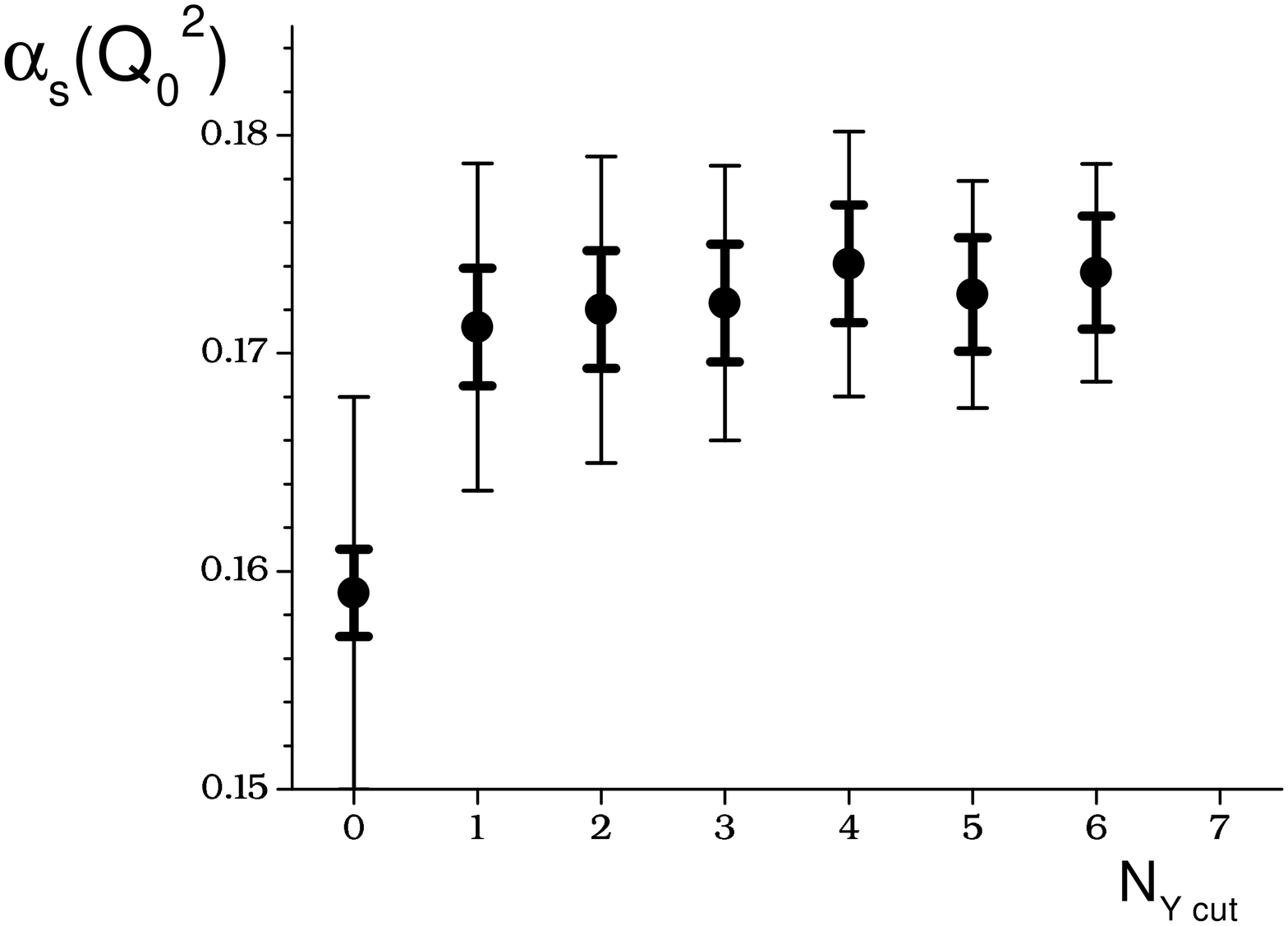} \\
\end{center}
\caption{
The study of systematics at different $Y_{cut}$ values
in the fits based on nonsinglet evolution.
The  QCD analysis of BCDMS ${}^{12}C, H_2, D_2$ data (nonsinglet case)
is given at
$x_{cut}=0.25$ and $Q_0^2=90$ GeV$^2$. 
The inner (outer) error-bars show statistical (systematic) errors.
}
 \label{fig:3}
\end{minipage}%
\hspace{0.04\textwidth}%
\begin{minipage}[t]{0.48\linewidth}
\begin{center}
\includegraphics[width=3.1in]{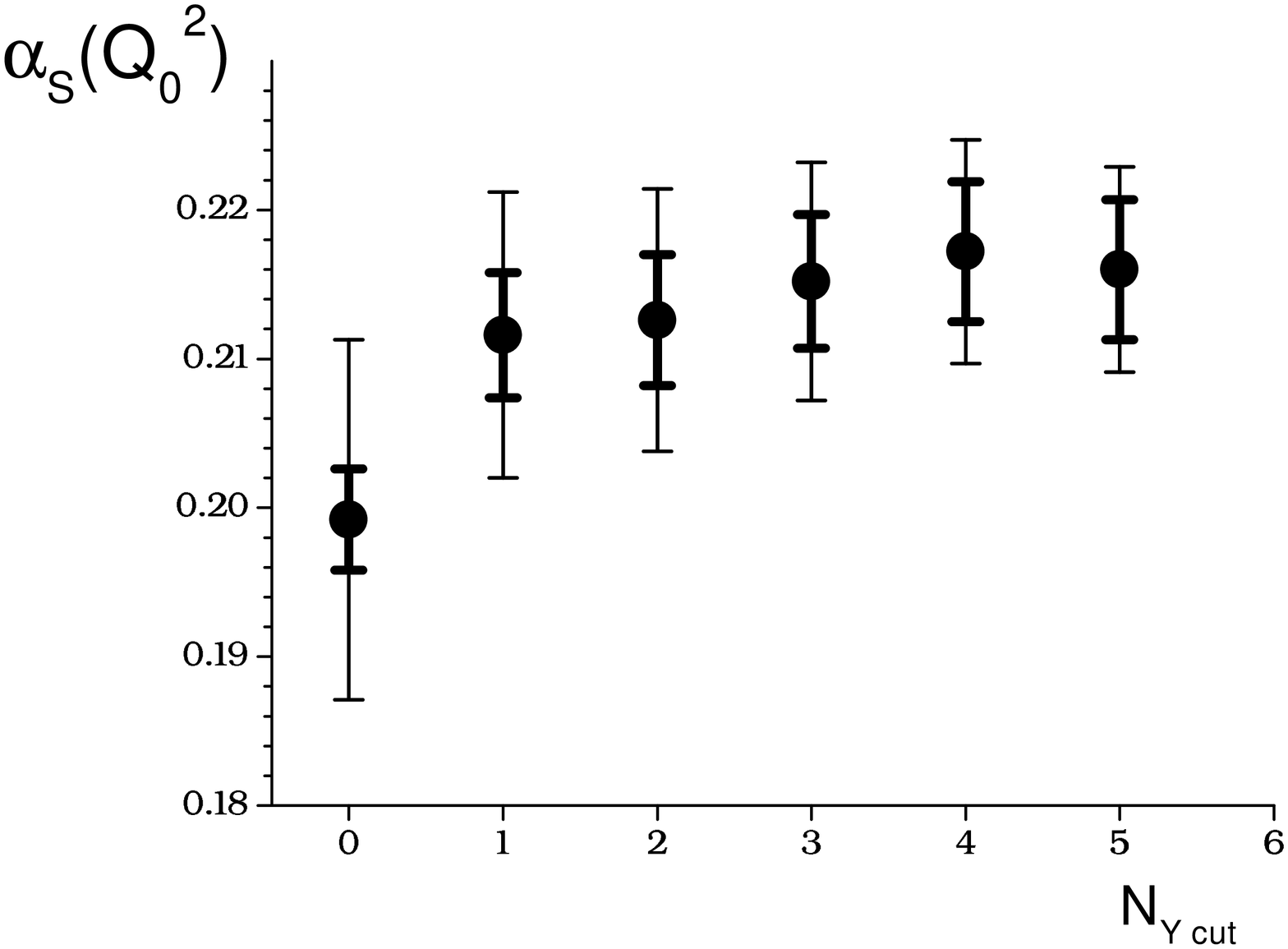} \\
\end{center}
 \caption{
The study of systematics at different $Y_{cut}$ values
in the fits based on combine singlet and nonsinglet evolution. 
All other notes are as in Fig. 1 with two exceptions:
no a $x_{cut}$ and  $Q_0^2=20$ GeV$^2$. 
Moreover, the points $N_{Ycut}=1,2,3,4,5$ correspond the values
$N=1,2,4,5,6$ in the Table 1.
}\label{fig:4}
\end{minipage}
\end{figure}

\vskip -0.5cm
The systematic errors for BCDMS data were given \cite{BCDMS1}
as multiplicative factors to be applied to $F_2(x,Q^2)$: $f_r, f_b, f_s, f_d$
and $f_h$ are the uncertainties due to spectrometer resolution, beam momentum,
calibration, spectrometer magnetic field calibration, detector inefficiencies 
and energy normalization, respectively.
For this study each experimental point of the undistorted set was multiplied
by a factor characterizing a given type
of uncertainties and a new (distorted) data set was fitted again
in agreement with our procedure considered in the previous section. The factors
($f_r, f_b, f_s, f_d, f_h$) were taken from papers \cite{BCDMS1}
(see CERN preprint versions in \cite{BCDMS1}).
The 
$\alpha_s$ values  
for the distorted
and undistorted sets of data are given in 
the Figs. 1 and 2 (for the cases of nonsinglet and complete evolutions,
respectively) together with the total systematic
error 
estimated in quadratures.

From 
the Figs. 1 and 2 we can see that the $\alpha_s$ values are obtained
for $N=1 \div 6$ of $Y_{cut3}$, $Y_{cut4}$ and $Y_{cut5}$ are very stable and
statistically consistent. The case $N=6$ of the Table 1 reduces the 
systematic error
in $\alpha_s$ by factor $1.8$ and increases the value of $\alpha_s$,
while increasing the statistical error on the 30\%.

After the cuts have been implemented 
(we use the set $N=6$ of the Table 1),
we have 590 (452) points (for the bounded 
$x$ range: $x \geq 0.25$).
Fitting them in agreement with the same procedure considered in the previous 
Section,
we obtain the following results:

from fits, based on nonsinglet evolution (i.e. when
$x \geq 0.25$):
\bea
\as(M_Z^2) &=& 0.1153 \pm 0.0013 ~\mbox{(stat)} 
\pm 0.0022 ~\mbox{(syst)} \pm 0.0012 ~\mbox{(norm)},
\nonumber
\eea

from fits, based on combined singlet and 
nonsinglet evolution:
\bea
\as(M_Z^2) &=& 0.1175 \pm 0.0014 ~\mbox{(stat)} 
\pm 0.0020 ~\mbox{(syst)} \pm 0.0011 ~\mbox{(norm)},
\label{ta}
\eea
where
hereafter the symbol 
``norm'' marks the 
error of normalization of experimental data.

The results are agree each other within considered errors. In Ref. 
\cite{KriKo} we have also analyzed the combine SLAC, NM and BFP data
and found 
good agreement with
(\ref{ta}). So, we have a possibility to fit together all the data.
It is the subject of the following subsection.

\subsection{ SLAC, BCDMS, NM and BFP data }
\label{subsec:revi1}
After these $Y$-cuts have been incorporated (with $N=6$) for BCDMS data, 
the full set of combine data is 1309 (797) points (for the bounded
$x$ range: $x \geq 0.25$).

To verify the range of applicability of perturbative QCD,
we analyze firstly the data without a contribution of twist-four terms,
i.e. when $F_2 = F_2^{pQCD}$. We do several fits using the cut 
$Q^2 \geq Q^2_{cut}$ and increase the value $Q^2_{cut}$ step by step.
We observe  good agreement of the fits with the data when 
$Q^2_{cut} \geq 10 \div 15$ GeV$^2$ (see the Figs. 3 and 4).
Later we add the twist-four corrections and fit the data with the
standard cut $Q^2 \geq 1$ GeV$^2$.
We have find very good agreement with the data. Moreover 
the predictions for $\asMZ$ in both above procedures 
are very similar (see the 
Figs. 3 and 4).
\begin{figure}[t]
\begin{minipage}[t]{0.48\linewidth}
\begin{center}
\includegraphics[width=3.0in]{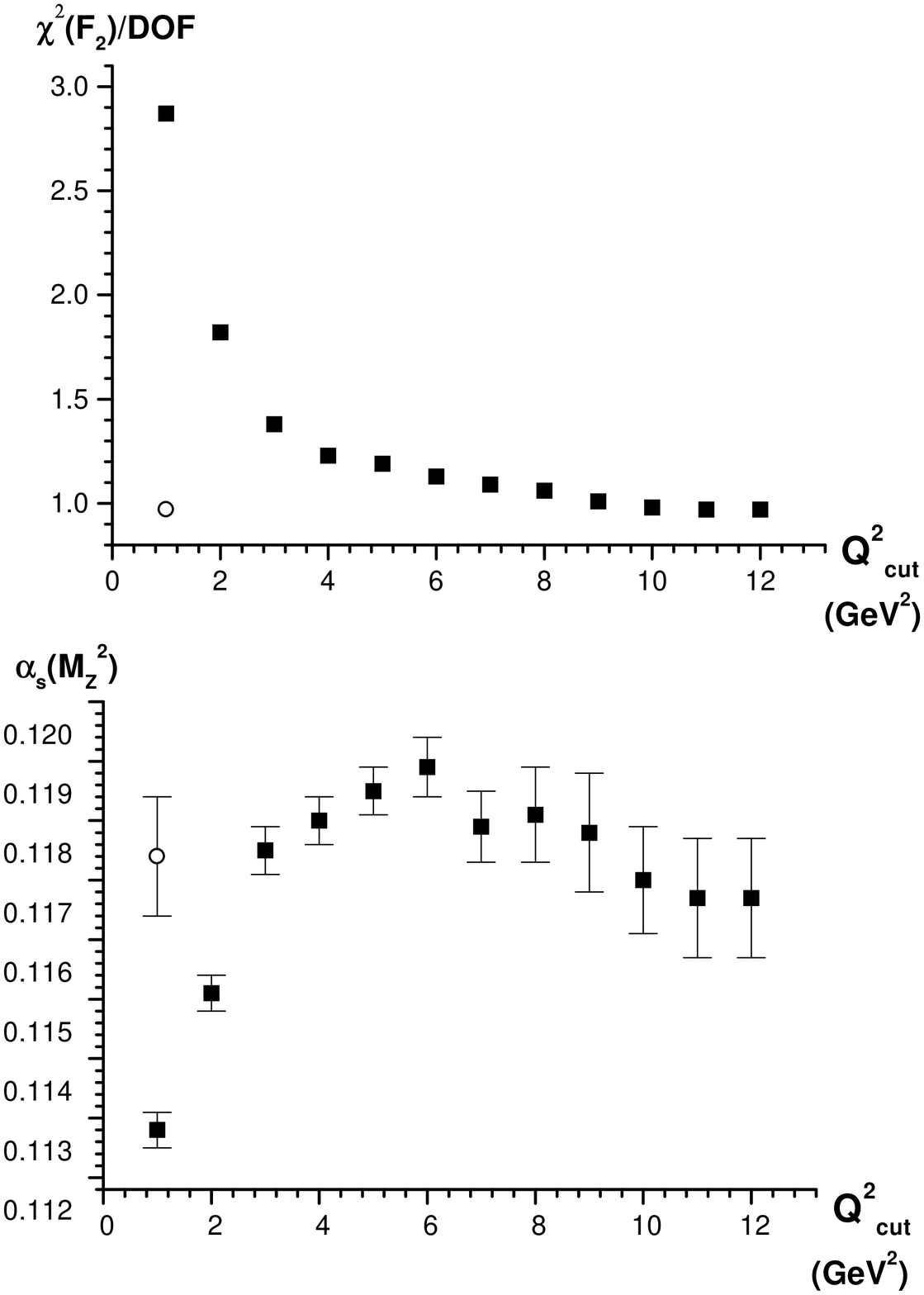} \\
\end{center}
\caption{
The values of $\asMZ$ and $\chi^2$ at different $Q^2$-values of data cuts
in the fits based on nonsinglet evolution.
The black (white) 
points show the 
analyses of data without (with) twist-four contributions.
Only statistical errors are shown.
}
 \label{fig:3}
\end{minipage}%
\hspace{0.04\textwidth}%
\begin{minipage}[t]{0.48\linewidth}
\begin{center}
\includegraphics[width=3.0in]{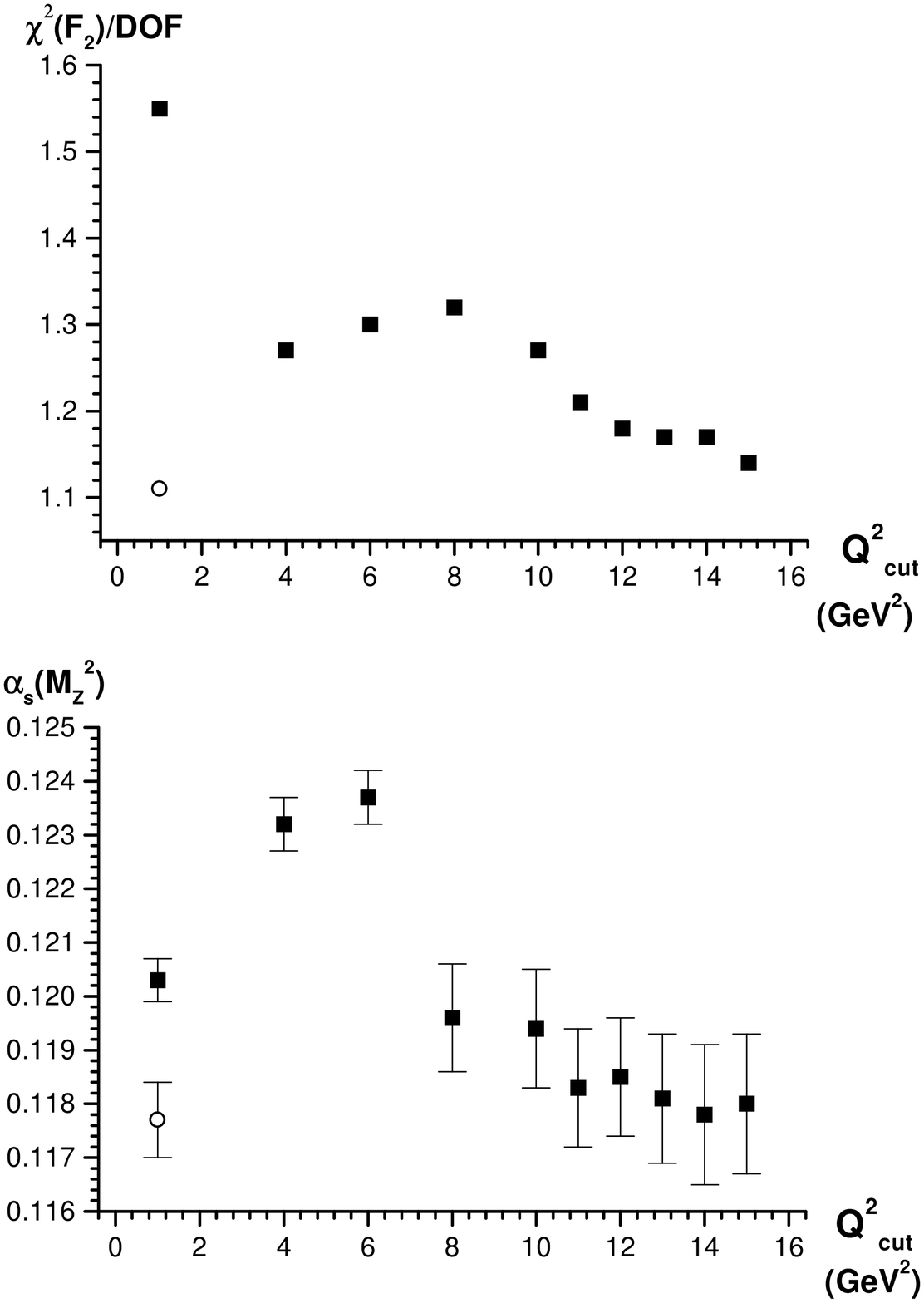} \\
\end{center}
 \caption{
The values of $\asMZ$ and $\chi^2$ at different $Q^2$-values of data cutes
in the fits based on combine singlet and nonsinglet evolution. 
All other notes are as in Fig. 3.
}\label{fig:4}
\end{minipage}
\end{figure}
The results of the fits are compiled in Summary (see Eqs. 
(\ref{re1n})-(\ref{re2s})).

\section{ Summary }

We
have demonstrated several steps of our study \cite{KriKo}
of the $Q^2$-evolution of DIS structure function $F_2$ fitting all
modern fixed target experimental data. 

From the fits we have obtained the value of the normalization 
$\asMZ$
of QCD coupling constant. First of all, we have reanalyzed the BCDMS data 
cutting the range with large systematic errors. As it is possible to see
in 
the Fig. 1, 
the value of $\asMZ$ rises strongly when
the cuts of systematics were incorporated. In another side, 
the value of $\asMZ$ does not dependent on the concrete type of the
cut within 
modern statistical errors.

We have found that at $Q^2 \geq 10 \div 15$ GeV$^2$ 
the formulae of pure perturbative
QCD (i.e. twist-two approximation together with target mass corrections)
are in good agreement with all data. 
\footnote{We note that at small $x$ values, the perturbative QCD
works well starting with $Q^2 = 1.5 \div 2$ GeV$^2$
and higher twist corrections are important only at very low $Q^2$:
$Q^2 \sim 0.5$ GeV$^2$ (see \cite{Q2evo,HT} and references therein).
As it is was observed in \cite{DoShi,bfklp} (see also discussions in
\cite{Q2evo,HT,BoAnd}) the good agreement between perturbative QCD and
experiment seems connect with large effective argument of coupling
constant at low $x$ range.}
 The 
results for  $\asMZ$ are very similar (see \cite{KriKo}) for the 
both types of analyses: ones, based on
nonsinglet evolution, and ones, based on combined singlet and 
nonsinglet evolution.
They have the following form:
\begin{itemize}
%
\item  from fits, based on nonsinglet evolution:
\bea 
\as(M_Z^2) &=& 0.1170 \pm 0.0009 ~\mbox{(stat)}
\pm 0.0019 ~\mbox{(syst)} \pm 0.0010 ~\mbox{(norm)}, \label{re1n} 
\eea
\item from fits, based on combined singlet and 
nonsinglet evolution:
\bea
\as(M_Z^2) &=& 0.1180 \pm 0.0013 ~\mbox{(stat)}
\pm 0.0021 ~\mbox{(syst)} \pm 0.0009 ~\mbox{(norm)}, 
\label{re1s}
\eea
\end{itemize}
\vskip -0.3cm

When we have added twist-four corrections, we have very good agreement
between QCD (i.e. first two coefficients of Wilson expansion)
and data starting already with $Q^2 = 1$ GeV$^2$, where the Wilson
expansion should begin
to be applicable.
The results for  $\asMZ$ coincide for the both types of analyses:
ones, based on
nonsinglet evolution, and ones, based on combined singlet and 
nonsinglet evolution.
They have the following form:
\begin{itemize}
\item  from fits, based on nonsinglet evolution:
\bea 
\as(M_Z^2) &=& 0.1174 \pm 0.0007 ~\mbox{(stat)}
\pm 0.0019 ~\mbox{(syst)} \pm 0.0010 ~\mbox{(norm)}, \label{re2n} 
\eea
\item from fits, based on combined singlet and 
nonsinglet evolution:
\bea
\as(M_Z^2) &=& 0.1177 \pm 0.0007 ~\mbox{(stat)}
\pm 0.0021 ~\mbox{(syst)} \pm 0.0009 ~\mbox{(norm)}, 
\label{re2s}
\eea
\end{itemize}
\vskip -0.3cm

Thus, there is very good agreement (see Eqs. (\ref{re1n}), (\ref{re1s}),
(\ref{re2n}) and (\ref{re2s}))
between results based on pure perturbative QCD at quite large $Q^2$ values
(i.e. at $Q^2 \geq 10 \div 15$ GeV$^2$) and the results based on 
first two twist terms
of Wilson expansion (at $Q^2 \geq 1$ GeV$^2$, 
where the Wilson expansion should  be applicable).

We would like to note that we have good agreement also with the analysis 
\cite{H1BCDMS} of
combined H1 and BCDMS data, which has been given by H1 Collaboration very 
recently. 
Our results for $\as(M_Z^2)$ are in good agreement also with 
the average value for coupling constant,
presented in the recent studies (see \cite{Al2000,NeVo,SaYnd,LEP}
and references therein) and in
famous Altarelli and Bethke reviews \cite{Breview}.

The last result (\ref{re2s}) based on all data with $Q^2 \geq 1$ GeV$^2$
can be considered as ``best
value'' for the coupling constant $\as(M_Z^2)$ coming in our analysis.

\vskip 0.2cm
{\bf Acknowledgments.}
~~The study is supported in part by 
the Heisenberg-Landau program.
Authors
 would like to express their sincerely thanks to the Organizing
  Committee of Vth Int. Conference ``Renormalization group 2002''
for the kind invitation, the financial support
 at  such remarkable Conferences, and 
 for fruitful discussions.
A.V.K. was supported in part by Alexander von Humboldt
fellowship and INTAS  grant N366.

\end{document}